\documentclass[twocolumn,english,reprint,prl]{revtex4-1}
\usepackage[T1]{fontenc}
\usepackage[latin9]{inputenc}
\usepackage{graphicx}

\makeatletter

\newcommand*\LyXThinSpace{\,\hspace{0pt}}

\usepackage{babel}

\usepackage{babel}

\makeatother

\usepackage{babel}
\begin{document}

\title{The REsonant Multi-Pulse Ionization injection }

\author{Paolo Tomassini$^{1}$, Sergio De Nicola$^{2}$, Luca Labate$^{1,3}$,
Pasquale Londrillo$^{4}$, Renato Fedele$^{2,5}$, Davide Terzani$^{2,5}$
and Leonida A. Gizzi$^{1,3}$ }

\affiliation{$^{1}$Intense Laser Irradiation Laboratory, INO-CNR, Pisa (Italy)}

\email{paolo.tomassini@ino.it}

\affiliation{$^{2}$Dip. Fisica Universita'  di Napoli Federico II (Italy)}

\affiliation{$^{3}$INFN, Sect. of Pisa, (Italy)}

\affiliation{$^{4}$INAF, Bologna (Italy)}

\affiliation{$^{5}$INFN, Sect. of Napoli (Italy)}
\begin{abstract}
The production of high-quality electron bunches in Laser Wake Field
Acceleration relies on the possibility to inject ultra-low emittance
bunches in the plasma wave. In this paper we present a new bunch injection
scheme in which electrons extracted by ionization are trapped by a
large-amplitude plasma wave driven by a train of resonant ultrashort
pulses. In the REsonant Multi-Pulse Ionization (REMPI) injection scheme,
the main portion of a single ultrashort (e.g Ti:Sa) laser system pulse
is temporally shaped as a sequence of resonant sub-pulses, while a
minor portion acts as an ionizing pulse. Simulations show that high-quality
electron bunches with normalized emittance as low as $0.08$ mm$\times$mrad
and $0.65\%$ energy spread can be obtained with a single present-day
100TW-class Ti:Sa laser system. 
\end{abstract}
\maketitle

\section{Introduction}

High-quality Laser Wake Field Accelerated (LWFA) electron bunches
are nowdays requested for several applications including Free Electron
Lasers \cite{FEL,FEL2,FEL3}, $X/\gamma$ radiation sources \cite{THOMSON,THOMSON2,THOMSON3,THOMSON4,THOMSON5}
and staged acceleration \cite{STAGING,STAGING9,STAGING2,STAGING3,STAGING4}.
While performances of self-injected bunches generated in the so-called
bubble-regime \cite{BUBBLE,BUBBLE4} continue to improve, other promising
injection schemes, including injection via density downramp \cite{TRANSITION,TRANSITION2,TRANSITION3,TRANSITION4,TRANSITION5},
colliding pulses injection \cite{COLLIDING,COLLIDING2,COLLIDING3}
and ionization injection \cite{IONIZATION,IONIZATION2,IONIZAIONNEW,IONIZATIONSHOCK,IONIZATION3,IONIZATION4,IONIZATION5,IONIZATION6-1,IONIZATION6},
are under active theoretical and experimental investigation.

Evolutions of the ionization injection, based on the use of two laser
pulses (either with the same or different wavelengths), were proposed
in \cite{twop0,TWOCOLOUR,TWOCOLOUR2,TWOCOLOUR3}. In the two-colour
ionization injection \cite{TWOCOLOUR} the main pulse that drives
the plasma wave has a long wavelength, five or ten micrometers, and
a large amplitude $a_{0}=eA/mc^{2}=8.5\cdot10^{-10}\sqrt{I\lambda^{2}}>1$,
being $I$ and $\lambda$ pulse intensity in $W/cm^{2}$ and wavelength
in $\mu m$. The second pulse (the ``ionization pulse'') possesses
a large electric field though its amplitude is low. This is achieved
by doubling the fundamental frequency of a Ti:Sa pulse. While the
main pulse cannot ionize the electrons in the external shells of the
contaminant species due to its large wavelength, the electric field
of the ioniziation pulse is large enough to generate newborn electrons
that will be trapped in the bucket. This opens the possibility of
using gas species with relatively low ionization potentials, thus
enabling separation of wake excitation from particle extraction and
trapping.

\begin{figure}
\includegraphics[bb=30bp 14bp 1546bp 673bp,clip,scale=0.16]{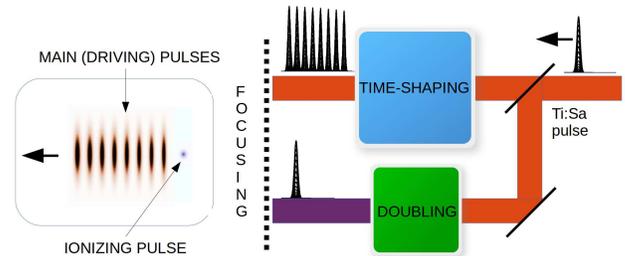}
\caption{Multi-Pulse ionization injection scheme. A small fraction of a single
Ti:Sa laser pulse is frequency doubled and will constitute the ionizing
pulse. The main portion of the pulse is temporally shaped as a train
of resonant pulses that will drive a large amplitude plasma wave. }
\end{figure}

Two colour ionization injection is therefore a flexible and efficient
scheme for high-quality electron bunch production. The main drawbacks
of the two colour ionization injection are the current lack of availability
of short (T<100 fs) 100TW-class laser systems operating at large ($\approx10\mu m$)
wavelength and lasers synchronization jitter issues. These limitation
make the two-colour scheme currently unpractical for application to
LWFA-based devices requiring high quality beams.

\begin{figure}
\includegraphics[scale=0.4]{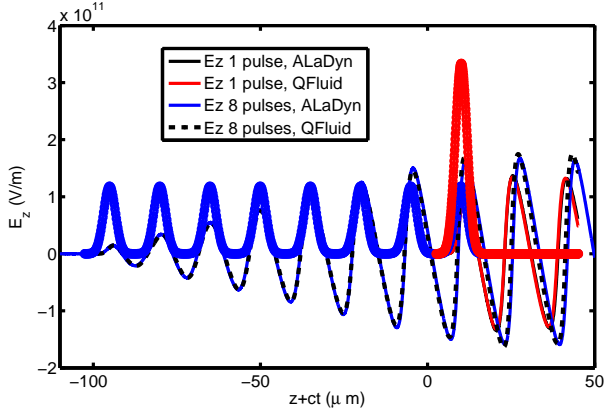}

\caption{Single-pulse vs eight-pulses train comparison. Pulses (moving through
the left) with duration of  $10 fs$  and waist size of 25 $\mu$
m are focused in a $n_{e}=5\times10^{18}$cm$^{-3}$ plasma. The single-pulse
(red thick line) with peak intensity of  $5.9\times 10^{18} W/cm^2$ 
drives a plasma wave whose maximum accelerating gradient is 20\% less
than that of the wave excited by the eight-pulses train having the
same delivered energy and intensity  $7.4\times 10^{17} W/cm^2$.
QFluid and PIC (ALaDyn 2D) simulation are in excellent agreement.}
\end{figure}

In this paper we propose a new injection configuration (we will refer
to it as REsonant Multi-Pulse Ionization injection, REMPI) that overcomes
these limitations and opens the way to a reliable generation of high
quality Laser Wakefield accelerators. The breakthrough of our REMPI
scheme is to replace the long wavelength driving pulse of the two-colour
scheme with a short wavelength, resonant multi-pulse laser driver.
Such a driver can be obtained via temporal shaping techniques from
a \textit{single}, linearly polarized, standard CPA laser pulse. A
minor fraction of the same pulse is frequency doubled (or tripled)
and used as ionizing pulse. Due to the resonant enhancement of the
ponderomotive force, a properly tuned train of pulses is capable of
driving amplitude waves larger than a single pulse with the same energy
\cite{MULTIPULSE,MULTIPULSE2} (see Fig. 2 where a comparison between
a single-pulse and an eight-pulses train is shown). Noticeably, since
the peak intensity of the driver is reduced by a factor equal to the
number of train pulses, it is also possible to match the conditions
of \textit{both} particle trapping and unsaturated ionization of the
active atoms level. Recently \cite{MP-EXP,MP-PRL} exciting experimental
results on the generation of such a time shaped pulses demonstrate
that a multi pulse scheme is obtainable with present day technology. 

In Sec II we set-up trapping conditions for electrons extracted in
a plasma wave driven by a resonant train of pulses. In Sec. III we
will discuss in details the process of electron extraction by a linearly
polarized ultraintense pulse. We carried on extensive numerical simulations
to evaluate applicability and robustness of the scheme. Among them,
in Sec. IV we will report on the simplest case of un-guided pulses
designed for a state-of-the-art 250TW Ti:Sa laser system. Finally,
Sec. VI is devoted to discussion of the results obtained by our simulations.
In the Appendices details on the ADK ionization model will be found,
along with a description of the hybrid fluid/kinetic QFluid code used
for the simulations.

\section{Trapping Conditions in REMPI}

To set conditions of particles trapping in the plasma wave we will
focus on a laser pulse configuration with a beam waist $w_{0}$ exceeding
the plasma wavelength $\lambda_{p}$, where the longitudinal ponderomotive
force dominates over radial wakefield force. In the $1D$ limit, the
Hamiltonian of a passive particle in the plasma wave is \cite{HAMILTIN1D}
$H=(1+u_{z}^{2})^{1/2}-\beta_{ph}u_{z}-\phi$ where $\beta_{ph}$
is wave phase velocity (transverse contribution to the Lorentz factor
has been neglected since relatively low values of the pulse amplitudes
will be considered here). The separatrix Hamiltonian $H_{s}$ decomposes
the phase space in a sequence of periodic buckets, so trapping of
newbonrn electrons occurs if the particle Hamiltonian satisfies $H\le H_{s}$,
i.e if

\begin{equation}
\phi_{e}\ge1-1/\gamma_{ph}+\phi_{min}
\end{equation}
being $\phi_{e}$ the normalized electrostatic potential at particle
extraction and $\phi_{min}$ the minimum potential. Eq. 1 clearly
states that trapping condition relies on wave phase velocity and on
wake electrostatic potential, i.e. on plasma density and normalized
electric field $E_{norm}=E_{z}/E_{0}$ solely, being $E_{0}=mc\omega_{p}/e$.
Exact solution of the fully nolinear wave equation in the 1D limit
gives us a relationship between the normalized electric field and
maximum/minimum potential \cite{HAMILTIN1D} $\phi_{max,min}=E_{norm}^{2}/2\pm\beta_{ph}\sqrt{(1+E_{norm}^{2}/2)^{2}-1}$.

\begin{figure}[h]
\includegraphics[clip]{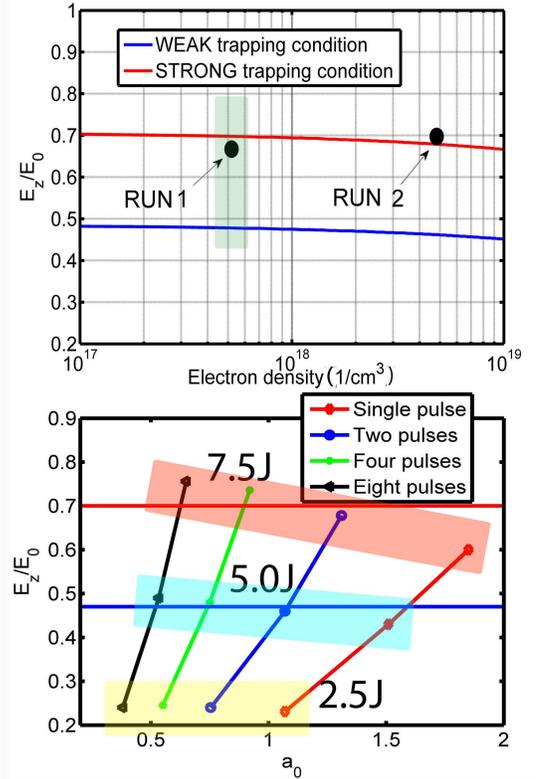} \caption{Trapping conditions. Blue lines: weak trapping threshold; red lines:
strong trapping condition. Top: trapping conditions in a 1D nonlinear
limit \textit{vs} plasma density from 1D analytical expression Eqq.
2 and 3. RUN 1,2 refer to the working points of the state-of-the-art
simulation (Sec. IV) and the simulation in Appendix II, respectively.
Bottom: scan of maximum accelerating normalized fields as in the RUN
1 setup ( $T=30\,fs,\ n_{e}=5\times10^{17}cm^{-3},\ w_{0}=45\,\mu m$)
as a function of pulse amplitude and the number of pulses in the train.
The cases of a single pulse and two, four and eight-pulses trains
with three different delivered energies of $2.5\,J$, $5.0\,J$ and
$7.5\,J$ have been considered.}
\end{figure}

Trapping starts when Eq. 1 holds, i.e. when electrons reach the end
of the bucket with the same speed as the wake phase speed ($v=\beta_{ph}c$).
Since these electrons will not be accelerated further, we will refer
to this condition as a ``weak trapping condition''

\begin{equation}
2\beta_{ph}\sqrt{(1+E_{norm}^{2}/2)^{2}-1}\ge1-1/\gamma_{ph}\:.
\end{equation}
Moreover, electrons that reach the speed of the wake before they experience
the maximum accelerating field will dephase in the early stage of
acceleration. As a consequence, a ``strong trapping condition''
can be introduced in such a way that electrons move with $v=\beta_{ph}c$
when they are in phase with the maximum longitudinal accelerating
field. In this case the potential at $E_{z}=E_{max}$ is null, so
we get

\begin{equation}
E_{norm}^{2}/2+\beta_{ph}\sqrt{(1+E_{norm}^{2}/2)^{2}-1}\ge1-1/\gamma_{ph}\:.
\end{equation}

Trapping analysis (see Fig. 3) reveals that efficient trapping occurs
in a nonlinear wave regime since $E_{norm}>0.5$, but well below longitudinal
wavebreaking for a cold nonrelativistic plasma ($E_{norm}<1$). Such
an analysis is confirmed by our simulations and it is useful to set
trapping threshold values for peak pulse amplitude $a_{0}$ in single
or multi-pulse schemes.

If a plasma density of $n_{e}=5\times10^{17}cm^{-3}$ is selected,
a matched set of parameters gives a pulse duration of $T=30$ fs FWHM,
with a minimum waist $w_{0}=45\,\mu$m (the same parameters set will
be used in the 250 TW state-of-the-art simulation, see below). A scan
of the maximum accelerating field versus pulse amplitude and the number
of pulses in the train is reported in Fig. 3. Three delivered energies
of $2.5\,$J, $5.0\,$J and $7.5\,$J have been considered and, for
any of them, a single-pulse, two, four and eight-pulses trains have
been simulated. As shown in Fig. 3 (bottom), for a fixed total delivered
laser energy, as the number of pulses in the train increases the maximum
accelerating gradient of the wake increases due to a resonance enhancement
of the wave. Moreover, from Fig. 3 (top and bottom) we can infer that
the weak-trapping threshold Eq.2 is reached with a single-pulse of
amplitude exceeding $a_{0}=1.6$, while in the case of a eight-pulses
train, weak-trapping threshold amplitude is reduced to $a_{0}=0.5$.

\section{Ionization dynamics in linear polarization}

Ultraintense laser pulses possess electric fields large enough to
make tunneling as the dominant ionization mechanism (i.e. Keldysh
parameter $\gamma_{K}=\sqrt{2U_{I}/mc^{2}}/a_{0}<<1$) so as the Ammosov-Delone-Krainov
(ADK) ionization rate \cite{ADK} can be assumed to evaluate electron
extraction from the initial level (see Appendix). Ionization potential
of $6^{th}$ electron from Nitrogen is $U_{I}^{6th}=552eV$ and efficient
extraction of $6^{th}$ electron of Nitrogen requires $a_{0}\approx1.7$
for a few tens of femtoseconds long pulses at $\lambda=0.8\,\mu m$.
On the other hand, Argon can be ionized from level $8^{th}$ to level
$9^{th}$ ($U_{I}^{9th}=422.5eV$) at a much lower intensity, being
$a_{0}\approx0.8$ and $a_{0}\approx0.4$ with $\lambda=0.8\mu m$
and $\lambda=0.4\,\mu m$, respectively.

We point out that a detailed description of ionization dynamics is
crucial not only to correctly estimate the number of bunch electrons
but (more importantly) to get a precise measure of the transverse
phase space covered by newborn electrons. In the linear polarization
case most of the electrons are ejected when the local electric field
is maximum, i.e. when the pulse potential $a_{e}$ is null. These
electrons will leave the pulse with a negligible quivering mean momentum
along the polarization axis $x$. If newborn electrons leave the atom
when electric field is not exactly at its maximum, a non null transverse
momentum $u_{x}=p_{x}/mc=-a_{e}$ is acquired, being $a_{e}$ the
local pulse potential at the extraction time. Moreover, ponderomotive
forces introduce an axisymmetrical contribution to particles transverse
momentum. Following \cite{IONIZMOM} we can write an expression for
the ${\it rms}$ momentum along the $x$ direction as a function of
the pulse amplitude \textit{envelope} at the extraction time $a_{0e}$
\begin{equation}
\sigma_{u_{x}}\cong\Delta\cdot a_{0e}=\sqrt{a_{0e}^{3}/a_{c}}
\end{equation}
where $a_{c}=0.107\left(U_{I}/U_{H}\right)^{3/2}\lambda$ is a critical
pulse amplitude and $\Delta=\sqrt{a_{0e}/a_{c}}$ (see Eqq. 7 and
10 in \cite{IONIZMOM}). Eq. 4 gives us an accurate estimate of the
minimum transverse momentum obtainable by the ionization process.

Trapping analysis with a standard \textit{single} pulse shows that
Nitrogen could be used in a simplified ionization injection (as suggested
in \cite{TWOCOLOUR}). Since efficient ionization threshold for $N^{6+}$
is $a_{0}\approx1.7$ for $\lambda=0.8\,\mu m$, a small interval
of $1.6<a_{0}<1.7$ for the pulse amplitude is suitable for both trapping
and ionization purposes. Such a simplified scheme could be useful
either for demonstration purposes or to obtain a controlled injection
for good-quality bunches without ultra-low emittance requirements.
A two-pulses driver is a far better choice since an optimal pulse
amplitude $1.1<a_{0}<1.3$ allows us to strongly inhibite driver pulses
ionization. Using Argon ($Ar^{8+}\rightarrow Ar^{9+}$) as a contaminant
instead of Nitrogen gives us a drastic reduction of transverse particle
momentum. Multi-pulse ionization injection with Argon requires trains
with at least four pulses since ionization level is saturated with
amplitude above $a_{0}=0.8$ at $\lambda=0.8\,\mu m$.

\section{State-of-the-art 250 TW simulation}

The reported simulation (RUN 1) of our REsonant Multi-Pulse Ionization
injection is based upon a linearly polarized Ti:Sa laser pulse that
is initially split into the ionizing pulse and the eight-pulses driver
train, each sub-pulse being $30\,fs$ FWHM in duration and delivering
$895\,$mJ of energy, with a maximum pulse amplitude $a_{0}=0.64$
and minimum waist size $w_{0}=45\,\mu$m. The uniform plasma electron
density is set to $n_{e}=5\times10^{17}$cm$^{-3}$ (plasma wavelength
is $\lambda_{p}=46.9\,\mu$m), obtained with a pure Argon pre-ionized
up to level $8^{th}$. The frequency doubled component ($\lambda_{ion}=0.4\,\mu$m)
with amplitude $a_{0,ion}=0.41$ and duration $T_{ion}=38\,$fs is
focused with a waist $w_{0,ion}=3.5\,\mu$m. The simulation (see Fig.
4) has been performed in a moving cylinder having a radius $4\times w_{0}$
with a resolution of $\lambda_{p}/100$ and $\lambda_{p}/200$ in
the radial and longitudinal coordinates, respectively. 

\begin{figure}
\includegraphics{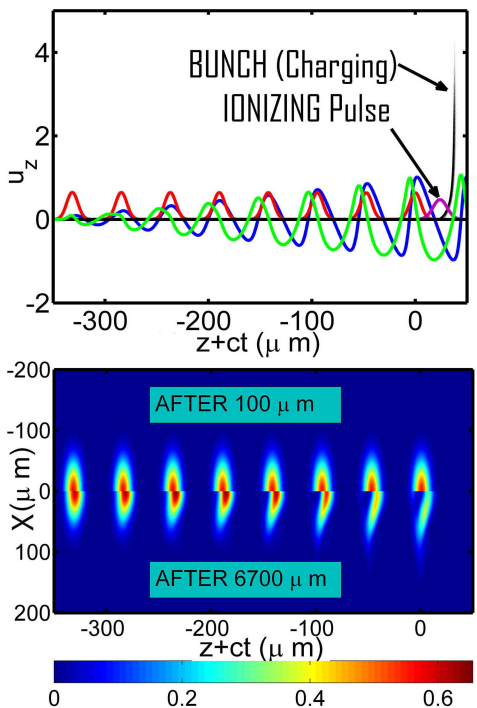} \caption{QFluid Snapshot after $100\,\mu m$ of propagation and after $6.5\,mm$.
Top: (after $100\,\mu m$) lineout of the pulses amplitudes (red/purple
lines), accelerating gradient (blue line) fluid longitudinal momentum
(green line) and extracted particle's longitudinal phase-space. Electrons
ejected by the driving pulse train don't comply with trapping conditions
and move as a (quasi) fluid. Bottom: laser pulse amplitude comparison
after $100\,\mu m$ (upper) and after $6.5\,mm$ (lower). }
\end{figure}

Electrons extracted in the bulk of the ionizing pulse move suddenly
backwards in the wake reaching the peak of the accelerating gradient
of relative intensity $E_{norm}=E_{z}/E_{0}=0.685$, i.e. very close
the the strong-trapping condition (see Fig. 3). Even though the driving
pulse sequence generates a marginal further ionization of $Ar^{8+}$
with a maximum percentage ionization of about $3\%$, such a dark
current will not be trapped in the wake (particles are extracted away
from the optimal extraction point of maximum potential $\phi_{max}$)
and so it will have no detrimental effect on beam quality. Moreover,
the short Rayleigh length $Z_{R}=\pi w_{0,ion}^{2}/\lambda_{ion}\approx100\,\mu$m
ensures a sudden truncation of beam charging that turns into a small
rms absolute energy spread $\Delta E\approx E_{norm}\times E_{0}\times Z_{R}\approx5\,$MeV
and extracted charge $Q=3.8\,$pC .

\begin{figure}[th]
\includegraphics{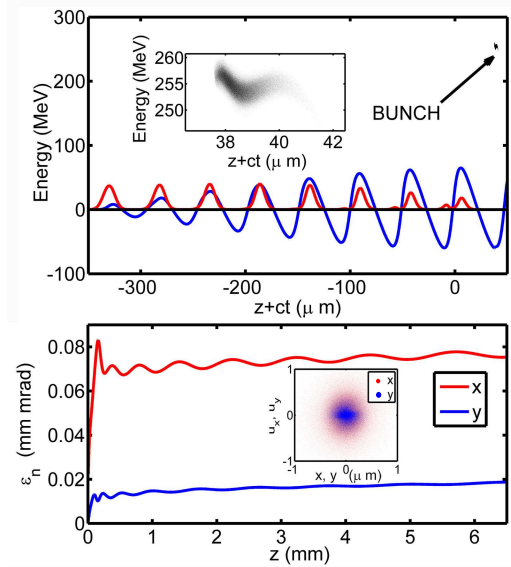} \caption{Bunch quality. Top: Final longitudinal position-energy distribution.
Blue and red lines represent the accelerating gradient and pulse amplitude
on axis on a.u., respectively. Inset: zoom of the longitudinal phase
space. Bottom: Normalized emittance in mm$\times$mrad as bunch moves
into the wake. Inset: final transverse phase space.}
\end{figure}

At the end of beam charging, i.e after about $200\,\mu$m of propagation,
the \textit{rms} bunch length is $0.56\,\mu$m and the transverse
normalized emittance is $\epsilon_{n,x}=0.070$ mm$\times$mrad in
the polarization direction $x$ and $\epsilon_{n,y}=0.016$ mm$\times$mrad
along the $y$ direction. Afterwards, the \textit{quasi-matched } beam
experiences dumped betatron oscillations with converging beam radius
of $0.4\,\mu$m that generate an emittance growth of about $10\%$
as simulation ends (see Fig. 5). 

Since in the weak nonlinear regime there's no electron density cavitation
as in the bubble regime, beam loading might be a serius limit for
beam quality. In the current working point, however, beam loading
is present but exerts a tiny perturbation (of about 1\%) of the longitudinal
field on the bunch core, as it is apparent in Fig. 6. We expect, therefore,
that the transverse asimmetry of the bunch ($\sigma(x)\approx2\sigma(y)$)
arising from the initial transverse momentum will generate asymmetric
beam loading effects but with very low amplitude. 

At the end of the $6.5\,mm$ long extraction/acceleration phase the
bunch has mean energy $265\,$MeV, with final normalized emittances
of $0.076$ mm$\times$mrad (x axis) and $0.018$ mm$\times$mrad
(y axis) , with an \textit{rms} energy spread $0.65\%$ and peak current
of about $1\,KA$. These extremely low values of emittance and energy
spread show that the proposed REsonant Multi-Pulse Ionization injection
scheme is ideal for the generation of very high quality accelerated
bunches. 

\begin{figure}
\includegraphics[scale=0.44]{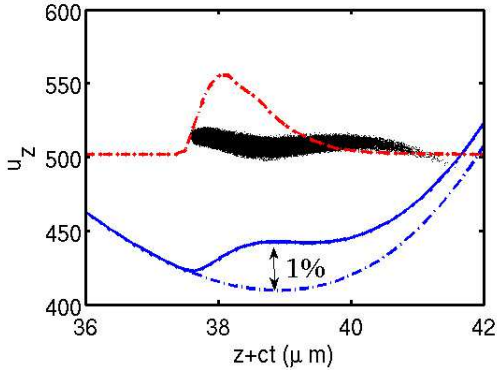}

\caption{Beam loading effect at the end of the simulation. The longitudinal
phase space of the beam is shown along with the (on axis) beam density
(red line), accelerating field (full blue line) and reference field
without beam-loading (dashed blue line). Beam loading makes a decrease
of the longitudinal field of about 1\% at most. }
\end{figure}

\section{Conclusions}

We propose a new ultra-low emittance LWFA injector scheme that uses
a Resonant train of pulses to drive plasma waves having amplitude
large enough to trap and accelerate electrons extracted by ionization.
The pulses train is obtained by temporal shaping of an ultrashort
pulse. Unlike two-colour ionization injection, a single laser system
(e.g Ti:Sa) can be therefore employed to both drive the plasma wave
and extract newborn electrons by ionization. Simulations consistently
show that the main processes, including extraction of electrons due
to the ionizing pulse, their trapping in the bucket and subsequent
acceleration can be controlled by tuning electron density and laser
intensity. Simulations also show a negligible contribution of spurious
electrons extracted directly by the driver pulses. Simulations carried
out in different plasma conditions show feasibility of the scheme
with state-of-the-art-lasers making REMPI suitable either for direct
interaction (e.g Thomson Scattering or FEL) or as ultra-low emittance
injector for GeV-scale energy boosting. 

Very recently J. Cowley et al. \cite{MP-PRL} reported on very encouraging
results about the feasibility of their time-shaping setup, with the
demonstration of efficient excitation of the plasma wave via Multi-Pulse
LWFA. The REMPI scheme could be tested with two pulses in the train
at first, with Nitrogen as a contaminant species. In order to obtain
very good-quality electron bunches, however, Argon should be preferred
and in this case more than four pulses in the train are necessary
as shown in Sec. II.

\section{Acknowledgments}

The research leading to these results has received funding from the
European Union's Horizon 2020 research and innovation programme under
Grant Agreement No 653782 - EuPRAXIA. Authors wish to thank CNAF-INFN
for access to computationl resources. Authors also acknowledge support
from Manuel Kirchen from Hamburg University for his help about the
FB-PIC code.

\section{Appendix I. ADK ionization rate}

In this paper we use the following formulation of the instantaneous
ADK ionization rate in the tunnelling regime \cite{ADK}

\begin{equation}
w_{ADK}(|m|)=C\times\rho_{ADK}^{n(|m|)}\times exp\left(-1/\rho_{ADK}\right)\:,\label{eq:ADK1}
\end{equation}
where  $n(|m|)=-2n^*+|m|+1$ ,  $C$  is a coefficient depending on
the atomic numbers and ionization energy  $U_I$  ( $U_H=13.6eV$ )

\begin{equation}
C=\frac{1}{4\pi}\left(\frac{U_{I}}{U_{H}}\right)^{3/2}3^{(2n^{*}-|m|-1)}\left[\frac{4e^{2}}{n^{*2}-l^{*2}}\right]^{n^{*}}\left[\frac{n^{*}-l^{*}}{n^{*}+l^{*}}\right]^{l+\frac{1}{2}},\label{eq:ADK2}
\end{equation}
and  $\rho_{ADK}={3/2}\left(E/E_{at}\right)\left(U_H/U_I\right)^{3/2}$,
being  $E_{at}=0.514\, TV/m$  and  $E$  the atomic and the local
electric fields, respectively. The effective quantum numbers are  $n^*=Z\sqrt{U_H/U_I}$ 
and  $l^*=n_0^*-1$, being  $n_0^*$  referred to the lower state
with the same  $l$. A critical electric field  $E_c={2/3}E_{at}\left(U_I/U_H\right)^{3/2}$,
giving a scale of a short-time scale ionization, can be introduced.
By expressing  $E/E_c$  using vector potentials we get  $a/a_c=\rho_{ADK}=9.37\left(U_H/U_I\right)^{3/2}a/\lambda$ 
which is nothing but the square of  $\Delta$  parameter in \cite{IONIZMOM}.

In the circularly polarized pulse case the electric field rotates
within each cycle still retaining the same intensity, so in the tunnelling
regime the mean-cycled ionization rate coincides with the instantaneous
rate of Eq. 5 
\begin{equation}
<w_{c}>=w_{ADK}\:.\label{eq:ADK3}
\end{equation}

In the linearly polarized pulse case the mean over a cycle can be
performed analytically after a taylor expansion of the leading exponential
term. The well known result (rewritten in our notation) is 

\begin{equation}
<w_{L}>=w_{ADK}(\rho_{ADK,0})\times(\frac{2}{\pi}\rho_{ADK,0})^{1/2}\:,\label{eq:ADK4}
\end{equation}
where $\rho_{ADK,0}$ is the peak value of $\rho_{ADK}=a/a_c$ within
the cycle. A numerical estimation of the mean-cycled rate confirms
the validity of Eq. 8 with errors below 4\% in the ionization rates,
for $\rho_{ADK}$ parameters in the range of interest for ionization
injection techniques (see Fig. 7).

\begin{figure}
\includegraphics[scale=0.44]{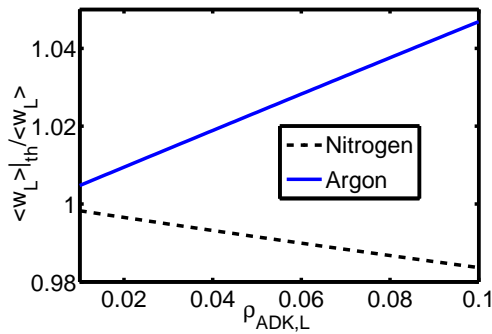}

\caption{Comparison between the numerical estimation of the mean-cycled ADK
rate and the widely used analytical result of Eq. 9 for $Ar^{9+}$
and $N^{6+}$ final states.}
\end{figure}

\section{Appendix II QFluid code}

QFluid is a cold-fluid/kinetic code in 2D \textit{cylindrical} coordinates,
employing plasma dynamics in a Quasi Static Approximation \cite{QSA}.
Electrons macroparticles move kinetically in a full 3D dynamics depicted
by the longitudinal $E_{z}$ and radial $E_{r}$ electric field, the
azimuthal magnetic field $B_{\phi}$ and ponderomotive forces due
to laser pulses. The main laser pulse train propagates following the
envelope evolution equation with the second time derivative included
\cite{ENVELOPE}, while the evolution of the ionization pulse follows
the Gaussian pulse evolution prescription. For our purposes, in the
absence of non-fluid plasma behavior, strong longitudinal background
gradients and radial anisotropies, QFluid returns the same results
of a 3D PIC code with much less demanding computation time/resources.
Particle extraction from atoms/ions is simulated with an ADK rate
including the mean over a pulse cycle, while newborn particles are
finally ejected with a random transverse momentum $u_{\perp}$, whose
rms value depends on the polarization of the pulses. For a linear
polarization (as for the ionizing pulse) we assigned $\sigma_{u_{x}}\cong\Delta\cdot a_{0e}=\sqrt{a_{0e}^{3}/a_{c}}$(see
Eq. 4), while for the circular polarization each extracted particle
is associated to a random extraction phase $\phi_{e}$ so as $u_{x}=a_{0e}\times cos(\phi_{e})$,
$u_{y}=a_{0e}\times sin(\phi_{e})$. Benchmark of QFluid with a multi-pulse
setup has been obtained in a nonlinear regime with ALaDyn \cite{ALADYN}(used
here in either a 3D with laser envelope configuration or a 2D slice
with a full-PIC pulse evolution) and with FB-PIC (quasi-3D PIC) \cite{FB-PIC}. 

The comparison of QFluid with FB-PIC is focused on a 2-pulses driver
scheme with Nitrogen as atomic species. The selected working point
consists of linearly polarized pulses of duration $T=30\:fs$, mimum
waist size $w_{0}=12\:\mu m$ and amplitude $a_{0}=1.2$ delayed by
a plasma wavelength ($\lambda_{p}=27\,\mu m$ with $n_{e}=1.5\times10^{18}cm^{-3}$).
FB-PIC simulations were performed with two azimuthal modes, i.e. possible
deviation from perfect azimuthal symmetry were included. 

\begin{figure}
\includegraphics[scale=0.42]{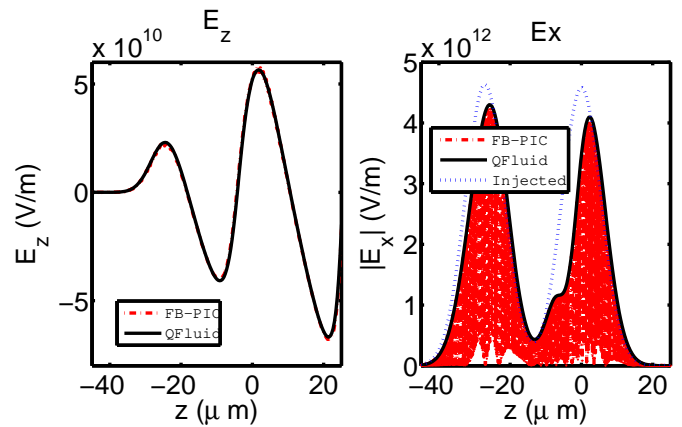}

\caption{FB-PIC vs QFluid in a two-pulses driver configuration with Nitrogen.
Snapshot after $700\,\mu m$ of propagation into a plasma with background
density of $n_{e}=1.5\times10^{18}cm^{-3}$. Left: longitudinal electric
field on axis. Right: pulse electric field (FB-PIC) and its amplitude
(QFluid). The injected pulse amplitude (blue dotted line) has been
shown for reference. }
 
\end{figure}

The comparison between FB-PIC and QFluid simulation (see Fig. 8) shows
a perfect superposition between the codes output, notwithstanding
the nontrivial evolution of the pulses due to both nonlinear effects
and the variation of the susceptivity due to the wake. 

The first QFluid and ALaDyn comparison shown here, has been focused
on an eight-pulses drivers train with Argon as atomic species, with
selected working point as the same as the state-of-the-art setup.
To fasten the 3D PIC simulation, ALaDyn has been equipped with an
envelope pulse solver. The Aladyn/envelope code implements a fully
3D PIC scheme for particle motion whereas the laser pulses are represented
by the envelope model proposed in \cite{ENV}.

\begin{figure}
\includegraphics[scale=0.4]{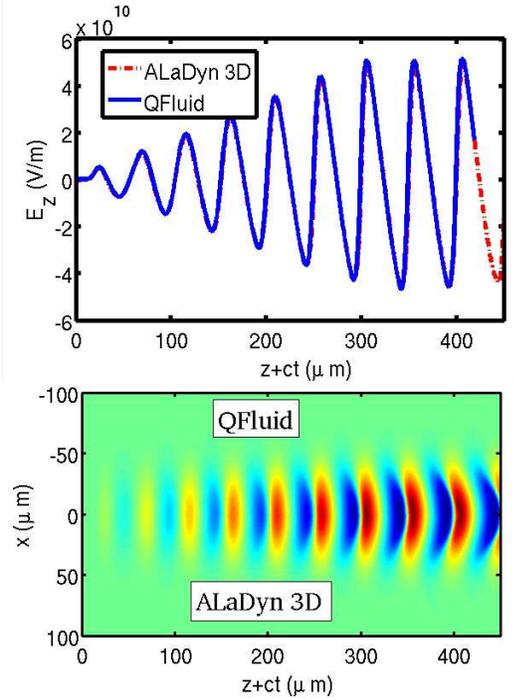}

\caption{ALaDyn vs QFluid in a eight-pulses setup with Argon (state-of-the-art
run parameters). Top: Snapshot of the on-axis longitudinal electric
field after $1\,mm$ of propagation. Bottom: radial maps of $E_{norm}=E_{z}/E_{0}$
for Qfluid (upper side) vs ALaDyn (lower side).}
\end{figure}

Once again (see Fig. 9), QFluid outcomes deviate at most of a few
percent from those of a 3D PIC (full 3D in this case). 

Finally, a full-PIC (not in envelope approximation) in 2D slice geometry
{\it vs} QFluid comparison, including the bunch extraction and trapping,
will be presented (RUN 2). To save computational time an high-density
setup has been simulated. A train of eight $10\,fs$ linearly polarized
Ti:Sa pulses impinge onto a preformed plasma of $Ar^{8+}$ with density
$5\times10^{18}$cm$^{-3}$. The driver pulse train has a waist size
$w_{0}=25\,\mu$m and an amplitude $a_{0}=0.589$, having a pulse
delay of a single plasma period $T_{p}=2\pi/\omega_{p}$. We use a
relatively large focal spot with $w_{0}>\lambda_{p}=14.8\,\mu$m,
so as to reduce the effects of the missing third dimension in the
PIC simulations. The frequency doubled ionizing pulse is injected
with a delay of $1.5\times T_{p}$ in the vicinity of its focus with
a waist $w_{0,inj}=3.5\,\mu$m and possesses a peak pulse amplitude
of $a_{0,inj}=0.41$. PIC simulations were performed with a $170\times150$
$\mu m^{2}$ box in the longitudinal and transverse directions with
a resolution of $\lambda/40$ and $\lambda/10$, respectively. QFluid
simulations were carried out in the same (cylindrical) box size with
resolution $\lambda_{p}/70$ and $\lambda_{p}/35$ in the longitudinal
and radial coordinates, respectively. 

\begin{figure}[tbph]
\includegraphics[clip,scale=0.4]{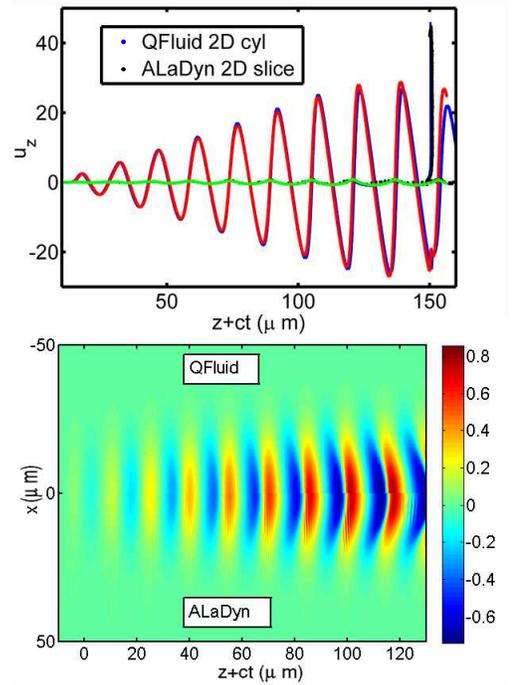} \caption{2D-slice ALaDyn and QFluid in RUN 2 setup. QFluid and ALaDyn PIC results
after $300\,\mu m$ of propagation. Top: (on-axis) ALaDyn phase space
of particles (black dots), QFluid phase space of particles (blue dots),
ALaDyn accelerating field (blue line, a.u.) and QFluid accelerating
field (red line, a.u.). The green line represents fluid momentum of
QFluid output. Bottom: Longitudinal electric field $E_{norm}=E_{z}/E_{0}$
from QFluid (upper) and ALaDyn (lower). }
\end{figure}

The final snapshot of both simulation, after $300\,\mu$m propagation
in the plasma is shown in Fig. 10, where the injected electron bunch
just at the end of the charging phase is visible (black and blue dots).
Due to the large ponderomotive forces (that scale as $a_{0,inj}^{2}/w_{0,inj}$,
see Eq. 23 in \cite{IONIZMOM}), bunch transverse \textit{rms} momentum
($0.26\,mc$ for QFluid and $0.27\,mc$ for ALaDyn, respectively)
shows an increase of about a factor of $2$ from the value expected
by Eq. 4. 

We finally stress that QFluid cannot face with the plasma exit of
the generated bunch since the Quasi Static Approximation requires
a steady plasma density within the box. A PIC code will be used in
a future work to simulate the plasma exit, too.

\end{document}